\documentclass[
a4paper,
aps,
prd,
preprint,
amsmath,
amssymb,
showpacs,
nofootinbib]{revtex4-1}
\usepackage[pdftex]{graphicx}
\usepackage{bm}

\begin{document}

\title{\boldmath Right-handed current with $CP$\/ violation 
       in the $b\to u$ transition}


\author{Tetsuya ENOMOTO}
\email{tetsuya@het.phys.sci.osaka-u.ac.jp}
\author{Minoru TANAKA}
\email{tanaka@phys.sci.osaka-u.ac.jp}
\affiliation{Department of Physics, Graduate School of Science, 
             Osaka University, Toyonaka, Osaka 560-0043, Japan}

\begin{abstract}
We show that the experimental data on $|V_{ub}|$ in various
$B$ meson decay modes suggest a possibility of $CP$\/-violating
right-handed current in the $b\to u$ transition. Its consequences
in $B\to\pi\pi,\,\rho\rho,\,DK$ are examined and compared with
experimental results in order to clarify possible signals of the $CP$ 
violation in these decay modes.
As a result, we find that the $CP$\/-violating right-handed current
is consistent with current experimental data. Its signal might be 
discovered by precise $CP$ measurements in future experiments.
\end{abstract}


\preprint{OU-HET-836}

\maketitle

\section{Introduction}
Left-handedness is the most interesting nature of the weak charged current.
The charged weak boson $W$ couples only to the left-handed fermions
in the standard model (SM), and this structure of the SM results in purely
left-handed charged currents. Another notable aspect of the weak charged 
current is the Cabibbo-Kobayashi-Maskawa (CKM) mixing in the quark sector
\cite{CABIBBO,KM}. Because of the CKM mixing the strength of the quark charged
current varies depending on the flavors involved in a transition, while 
the left-handedness remains universal apart from tiny radiative corrections.

New physics (NP) beyond the SM may violate the universality of the charged 
current.
In particular, NP effects are expected to be significant for
CKM suppressed transitions if the supposed NP has a different
flavor structure from the SM. Among several CKM suppressed channels,
the $b\to u$ transition is most likely to be affected by NP in this sense.
Moreover, we can examine it with the competent data of $B$ factory experiments.

It was pointed out that a mixture of right-handed current (RHC)
in the $b\to u$ charged current explained the discrepancy among 
magnitudes of the relevant CKM matrix element, $|V_{ub}|$,
determined by various $B$ decay modes \cite{CN2008,CRIVELLIN2010,BGI2011}. 
Recently, Belle collaboration updated the data on the pure tauonic $B$ decay, 
$B^-\to\tau\bar\nu$ 
\footnote{Charge-conjugation modes are implied unless otherwise stated.}
\cite{BELLETAU2013}, and the new result seems more consistent with 
the SM than the previous one \cite{CP2014}.

In this work, we revisit the possibility of RHC in the $b\to u$ transition 
taking the new Belle data into account.
In Sec.~\ref{Sec:Vub}, we introduce the $b\to u$ RHC and
explain how it affects the $|V_{ub}|$ determination in leptonic and 
semileptonic $B$ decays. 
Then, we show that the present experimental data suggest a sizable 
$CP$\/-violating (CPV) RHC in this transition. According to this observation, 
we study possible CPV signals in hadronic $B$ decays, $B\to\pi\pi$, 
$B\to\rho\rho$ and $B\to DK$, in the presence of the $b\to u$ RHC in
Sec.~\ref{CPV}. We find that new CPV signals absent in the SM arise in 
direct $CP$ asymmetries and the measurement of angles of the unitarity 
triangle. Comparing these possible signals of the RHC with the relevant 
experimental data, we show that the $b\to u$ CPV RHC is a viable NP scenario.
We also compare the $b\to u$ RHC induced by squark mixings in the minimal 
supersymmetric standard model (MSSM) to the experimental constraint obtained
in Sec.~\ref{Sec:Vub}. Our conclusion is given in Sec.~\ref{CONCLUSION}.

\section{\label{Sec:Vub}\boldmath Effects of the right-handed current 
         in $|V_{ub}|$ determination}
A set of right-handed quark charged currents appears in the SM
once we introduce higher dimensional operators. 
The gauge-invariant effective lagrangian containing the lowest dimensional
operator responsible for the $b\to u$ transition is expressed by
\begin{equation}
\mathcal{L}_R=\frac{C_R}{\Lambda^2}\,\tilde\phi^\dagger iD_\mu\phi\,
              \bar u_R \gamma^\mu b_R
              +\text{h.c.}\,,
\end{equation}
where $\phi$ is the Higgs doublet, $\tilde\phi=i\sigma_2\phi^*$, 
$\Lambda$ represents the energy scale of NP, and 
$C_R$ is a dimensionless constant that depends on the details of NP. 
This lagrangian leads to the desired right-handed interaction
as a result of the electroweak symmetry breaking. 

Combined with the ordinary left-handed interaction, 
the $b\to u$ transition is described by
the following modified charged current lagrangian:
\begin{equation}
\label{Eq:LCC}
\mathcal{L}_\text{CC}
= -\frac{g}{\sqrt{2}}\bar u \gamma^\mu(V_{ub}^L P_L+V_{ub}^R P_R)b\,W_\mu^+
   +\text{h.c.}\,,
\end{equation}
where $g$ is the SU(2) gauge coupling, $P_{L(R)}=(1\mp\gamma_5)/2$,
and $V_{ub}^L$ denotes the left-handed CKM matrix element.
The effective right-handed CKM matrix element $V_{ub}^R$ is given by
\begin{equation}
V_{ub}^R=C_Rv^2/2\Lambda^2\sim 3\times 10^{-3}\,C_R
         \left(\frac{3\ \text{TeV}}{\Lambda}\right)^2\,,
\end{equation}
where $v\simeq 246\ \text{GeV}$ denotes the vacuum expectation value 
of $\phi$.
The known magnitude of the $b\to u$ charged current corresponds to 
$|V_{ub}^L|\sim 3.6\times 10^{-3}$ and the right-handed
component induced by NP at the TeV scale is potentially comparable. 

In the rest of this section, we treat $V_{ub}^L$ and $V_{ub}^R$ as
complex parameters and determine their values using experimental
data of leptonic and semileptonic $B$ meson decay modes 
along with the unitarity of the left-handed CKM matrix $V^L$ assuming that 
no SM interactions besides the RHC in Eq.~\eqref{Eq:LCC} are affected by NP. 
We emphasize that the nonvanishing relative phase between $V_{ub}^L$ and 
$V_{ub}^R$ is a new CPV degree of freedom.

\subsection{\boldmath $B\to\tau\bar\nu$}
The pure tauonic $B$ decay is a theoretically clean mode to determine 
$|V_{ub}|$ provided that the $B$ meson decay constant $f_B$ is known 
accurately enough. The decay rate is written as
\begin{equation}\label{Eq:BtoPiRate}
\Gamma(B^-\to\tau^-\bar\nu)=
 \frac{G_F^2 m_B m_\tau^2}{8\pi}\left(1-\frac{m_\tau^2}{m_B^2}\right)
 f_B^2|V_{ub}^\text{exp}|^2\,,
\end{equation}
where $|V_{ub}^\text{exp}|$ is the effective CKM matrix element that is
determined by experiments assuming the SM. Since the axial-vector current 
contributes to this decay mode, one finds 
$|V_{ub}^\text{exp}|=|V_{ub}^L-V_{ub}^R|$ in the presence of the RHC
in Eq.~\eqref{Eq:LCC}.

The present world average of the branching ratio including 
the updated Belle data is provided by the heavy flavor averaging group
(HFAG) \cite{HFAG} as
$\text{Br}(B\to\tau\bar\nu)=(1.14\pm 0.22)\times 10^{-4}$. 
Using this value and $f_B=190.5\pm 4.2\ \text{MeV}$ \cite{FLAG}, we obtain
\begin{equation}
|V_{ub}^\text{exp}|=|V_{ub}^L-V_{ub}^R|=(4.22\pm 0.42)\times 10^{-3}\,.
\label{Eq:Btau}
\end{equation}

\subsection{\boldmath $B\to\pi\ell\bar\nu$}
The axial-vector part in Eq.~\eqref{Eq:LCC} does not contribute to 
the process $B\to\pi\ell\bar\nu$ owing to the parity invariance of 
the strong interaction. The contribution of the vector part leads to 
the following differential decay rate,
\begin{equation}
\frac{d\Gamma}{dq^2}=\frac{G_F^2}{192\pi^2 m_B^3}\lambda^{3/2}(q^2)f_+^2(q^2)
                     |V_{ub}^\text{exp}|^2\,,
\label{Eq:BpiRate}
\end{equation}
where $q^\mu$ represents the momentum transfer, 
$\lambda(q^2)=q^4-2 q^2(m_B^2+m_\pi^2)+(m_B^2-m_\pi^2)^2$, and
$|V_{ub}^\text{exp}|=|V_{ub}^L+V_{ub}^R|$, which reduces to the ordinary
CKM matrix element in the absence of the right-handed interaction. 
The hadronic form factor $f_+(q^2)$ plays a crucial role in 
the determination of $|V_{ub}^\text{exp}|$. We employ the one evaluated
using the light-cone sum rule (LCSR) \cite{BZ2005P} and its explicit form 
is given in Appendix \ref{Ap:HFFP} for completeness. The mass of 
the charged lepton is neglected in Eq.~\eqref{Eq:BpiRate}.

Evaluating the branching ratio with Eqs.~\eqref{Eq:BpiRate} and 
\eqref{Eq:BZpi},
and comparing it to the experimental result \cite{HFAG}, 
$\text{Br}(B\to\pi\ell\bar\nu,0<q^2<16\ \text{GeV}^2)
 =(1.06\pm 0.04)\times 10^{-4}$, we obtain
\begin{equation}
|V_{ub}^\text{exp}|=|V_{ub}^L+V_{ub}^R|=(3.58\pm 0.47)\times 10^{-3}\,,
\label{Eq:Bpi}
\end{equation}
where the error includes 13\% theoretical uncertainty in the magnitude of
$f_+(q^2)$ \cite{BZ2005P} as well as the experimental errors.

\subsection{\boldmath $B\to(\rho,\omega)\ell\bar\nu$}
Both the vector and axial vector currents give rise to the decay
$B\to V\ell\bar\nu$, where $V$ denotes a vector meson and we consider
two modes with $V=\rho,\omega$. The differential decay rate is 
given as
\begin{equation}\label{Eq:BVRate}
\frac{d\Gamma}{dq^2}
=\frac{G_F^2|\bm{p}_V|q^2}{96\pi^3 c_V^2 m_B^2}
 \sum_{\lambda=\pm,0}|H_\lambda|^2\,,
\end{equation}
where $\bm{p}_V$ represents the three momentum of the vector meson in 
the rest frame of the $B$ meson, $c_V=\sqrt{2}$ for $V=\rho^0,\omega$ 
considered below, and the helicity amplitudes $H_{\pm,0}$ are expressed in 
terms of hadronic form factors $V(q^2)$ and $A_{1,2}(q^2)$:
\begin{align}
\label{Eq:Hpm}
H_\pm&=(V^L_{ub}-V^R_{ub})(m_B +m_V)A_1(q^2)\mp
       (V^L_{ub}+V^R_{ub})\frac{2m_B|\bm{p}_V|}{m_B+m_V}V(q^2)\,,\\
\label{Eq:Hz}
H_0&=(V^L_{ub}-V^R_{ub})\frac{m_B +m_V}{2m_V\sqrt{q^2}}
     \left\{(m_B^2 -m_V^2-q^2)A_1(q^2)
            -\frac{4m_B^2 |\bm{p}_V|^2}{(m_B+m_V)^2}A_2(q^2)\right\}\,.
\end{align}
We use LCSR form factors \cite{BZ2005V} and they are explicitly given
in Appendix \ref{Ap:HFFV}.

We can experimentally determine the magnitude of the $b\to u$ charged 
current, represented by $|V_{ub}^\text{exp}|$,
from the branching ratio of $B\to V\ell\bar\nu$. 
Integrating the differential rate in Eq.~\eqref{Eq:BVRate} over 
the phase space, we find 
\begin{equation}\label{Eq:BV}
|V_{ub}^\text{exp}|=
 |V_{ub}^L|\sqrt{1+a_V\mathrm{Re}\left(\frac{V_{ub}^R}{V_{ub}^L}\right)
                  +\left|\frac{V_{ub}^R}{V_{ub}^L}\right|^2}\,,
\end{equation}
where $a_V=-1.18(-1.25)$ for $V=\rho(\omega)$. In our numerical analysis, 
we employ the following experimental results \cite{BELLEVub2013}:
\begin{equation}\label{Eq:Brho}
|V_{ub}^\text{exp}|=(3.56\pm0.48)\times 10^{-3}\,,
\end{equation}
for $B\to\rho^0\ell\bar\nu$ and
\begin{equation}\label{Eq:Bomega}
|V_{ub}^\text{exp}|=(3.08\pm0.49)\times 10^{-3}\,,
\end{equation}
for $B\to\omega\ell\bar\nu$, where the errors include theoretical
uncertainties in the form factors. Since significant part of theoretical
uncertainties is involved in these errors, neither the theoretical
uncertainty in $a_V$ nor the effect of experimental cut in the phase space
integration performed to obtain Eq.~\eqref{Eq:BV} is taken into account 
in the present analysis. We note that a precise study of decay distribution 
in these modes with copious data at the SuperKEKB/Belle II experiment
may provide an improved method to probe the RHC \cite{BLT2014}.

\subsection{\boldmath $B\to X_u\ell\bar\nu$}
Both the left- and right- handed currents contribute to the inclusive
$b\to u$ semileptonic process. However the interference between them is
strongly suppressed because of the small mass of the up quark. 
Thus, the decay rate is proportional to 
$|V_{ub}^\text{exp}|^2=|V_{ub}^L|^2+|V_{ub}^R|^2$.
In our numerical analysis, we use the following result of GGOU 
method \cite{GGOU} given by HFAG \cite{HFAG}:
\begin{equation}
|V_{ub}^\text{exp}|=\sqrt{|V_{ub}^L|^2+|V_{ub}^R|^2}
=(4.39\pm 0.31)\times 10^{-3}\,,
\label{Eq:Inc}
\end{equation}
where the statistic and systematic errors are linearly added 
in order to take the sizable method dependence into account. 

\subsection{\label{SUBSEC:UT}Unitarity triangle}
Since the above five processes of direct $|V_{ub}|$ measurement
are described by two independent quantities, $|V_{ub}^L|^2+|V_{ub}^R|^2$
and $\mathrm{Re}(V_{ub}^L V_{ub}^{R*})$, they are insufficient to
determine the absolute values of $V_{ub}^L$ and $V_{ub}^R$,
and their relative phase. In order to extract more information on
$V_{ub}^L$ and $V_{ub}^R$, we utilize the unitarity of the CKM matrix
$V^L$ assuming the validity of the SM except for the $V_{ub}^R$ term 
in Eq.~\eqref{Eq:LCC} as stated previously.

\begin{figure}
 \centering
 \includegraphics[width=25em]{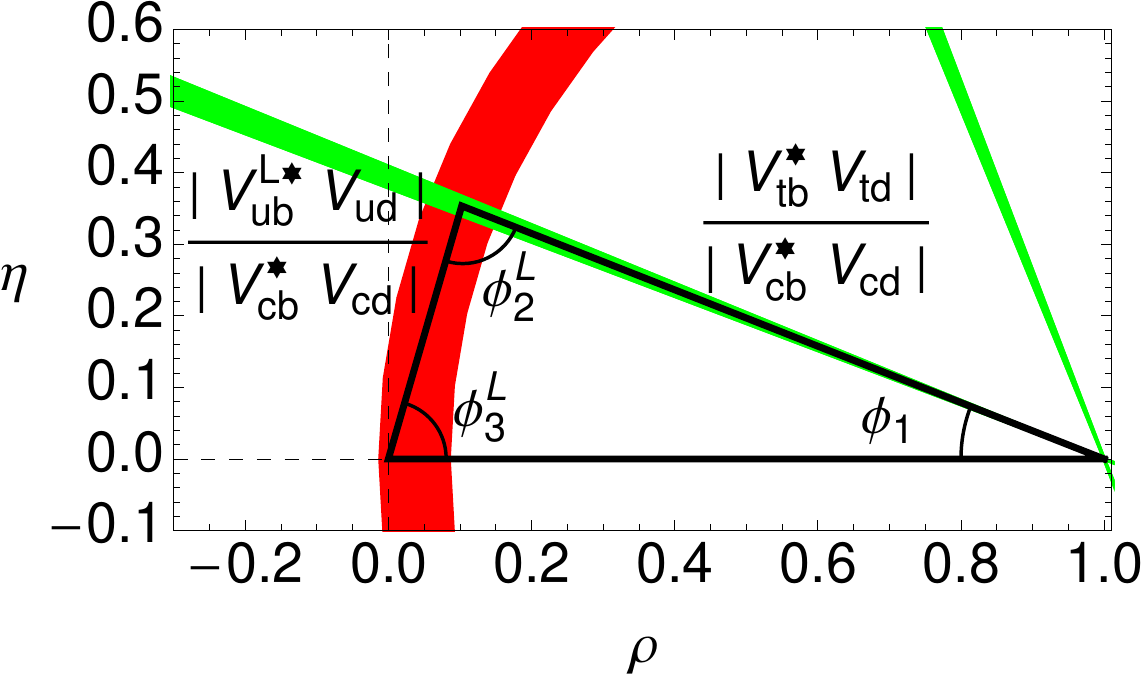}
 \caption{\label{Fig:RHOETA}
  Unitarity triangle (solid lines). 
  Experimental constrains from
  the $B$--$\bar B$ mixings and $CP$ violation in $b\to c\bar cs$ decays
  are also indicated by the red arc and the green narrow sectors respectively.}
\end{figure}

The unitarity of $V^L$ is conveniently represented by the unitarity
triangle in Fig.~\ref{Fig:RHOETA}, where Wolfenstein 
parameterization \cite{WP,PDG} is introduced. 
Measuring $|V_{td}V_{tb}^*|$ by the $B$--$\bar B$ mixings and
$\phi_1$ (or $\beta$) with $CP$ violation in $b\to c\bar cs$ decays,  
together with results of kaon and $b\to c$ semileptonic decays
that give $\lambda$ and $A$ in Wolfenstein parameterization,
we can indirectly determine the magnitude and phase of $V_{ub}^L$.

The mass difference in the $B_d$ meson system due to the $B_d$--$\bar B_d$
mixing, denoted by $\Delta m_{B_d}$, is dominated by the top quark loop 
in the SM and proportional to $|V_{td}V_{tb}^*|^2$; and 
similarly in the $B_s$ system, 
$\Delta m_{B_s} \propto |V_{ts}V_{tb}^*|^2$. 
Theoretical uncertainties of the relevant hadronic matrix elements
are reduced by taking the ratio of $\Delta m_{B_d}$ and $\Delta m_{B_s}$
owing to the SU(3) flavor symmetry:
\begin{equation}
\frac{\Delta m_{B_d}}{\Delta m_{B_s}}
=\frac{m_{B_d}}{m_{B_s}}\xi^{-2}\left\vert\frac{V_{td}V_{tb}^*}
                                               {V_{ts}V_{tb}^*}
                                  \right\vert^2
=\frac{m_{B_d}}{m_{B_s}}\xi^{-2}\lambda^2\left\{(1-\rho)^2+\eta^2\right\}\,,
\end{equation}
where $\xi=1.268\pm 0.063$ \cite{FLAG} represents the SU(3) breaking effect.
Thus, we determine $|V_{td}V_{tb}^*|$ from the present experimental data, 
$\Delta m_{B_d}=0.510\pm 0.003\ \text{ps}^{-1}$ and
$\Delta m_{B_s}=17.761\pm 0.022\ \text{ps}^{-1}$ \cite{HFAG}.
The result is shown as the red arc in Fig.~\ref{Fig:RHOETA}.

The time-dependent $CP$ asymmetries in $b\to c\bar c s$ processes such as 
$B\to J/\psi K_S$ give $\sin2\phi_1$ with small theoretical uncertainty 
in the SM. This argument does not change in the presence of the $b\to u$ 
RHC. The combined experimental data $\sin 2\phi_1=0.682\pm 0.019$ 
\cite{HFAG} gives $\phi_1$ with a four-fold ambiguity. 
It turns out that only the solution favored in the SM, as depicted in 
Fig.~\ref{Fig:RHOETA}, is consistent with the RHC. 

Consequently, one of the apices of the unitarity triangle $(\rho,\eta)$ 
is uniquely determined (with errors) and $V_{ub}^L=\lambda^3 A (\rho-i\eta)$
is evaluated as
\begin{equation}\label{Eq:UT}
|V_{ub}^L|=(3.43\pm 0.16)\times 10^{-3}\,,\ 
\phi_3^L=\arg V_{ub}^{L*}=73.8^\circ\pm 7.5^\circ\,,
\end{equation}
where $\lambda=0.225$ and $A=0.823$ are used \cite{CKMF}.

\subsection{Combined result}
\begin{figure}
 \centering
 \includegraphics[width=25em]{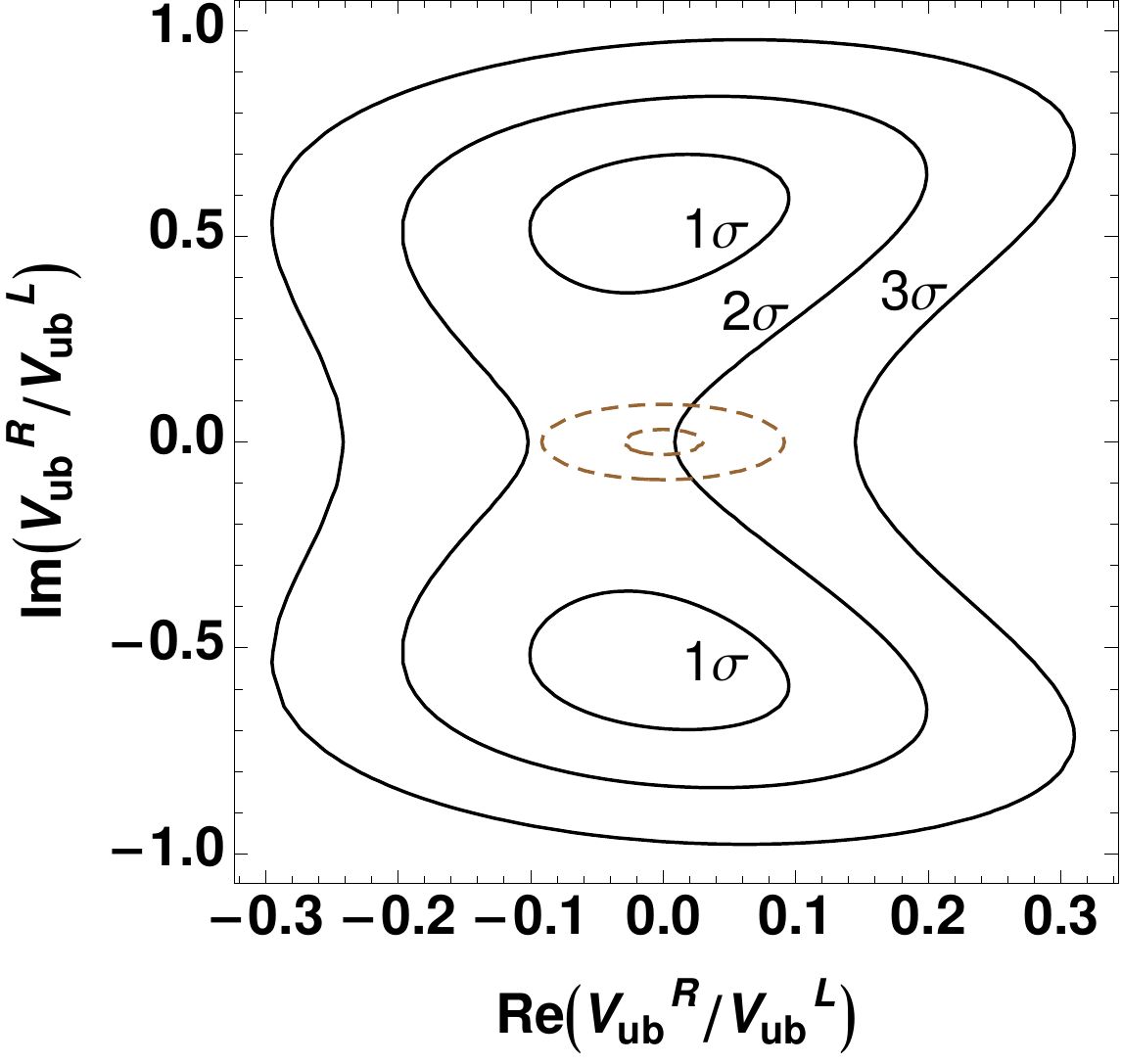}
 \caption{\label{Fig:VR}
  Allowed region of $V_{ub}^R/V_{ub}^L$ (solid lines). 
  The constraints from $|V_{ub}|$ and the unitarity triangle are combined. 
  Prediction of MSSM is also indicated by dashed lines. 
  (See Sec.~\ref{MSSM}.)}
\end{figure}
Combining the results in Eqs.~\eqref{Eq:Btau}, \eqref{Eq:Bpi}, \eqref{Eq:BV},
\eqref{Eq:Brho}, \eqref{Eq:Bomega}, \eqref{Eq:Inc}
and \eqref{Eq:UT}, we obtain a constraint on $V_{ub}^R$. 
The numerical result is presented in Fig.~\ref{Fig:VR}, 
where $1\sigma$, $2\sigma$ and $3\sigma$ allowed regions are indicated
by solid lines. The best fit is given by
\begin{equation}
\mathrm{Re}\left(\frac{V_{ub}^R}{V_{ub}^L}\right)=-4.21\times 10^{-3}\,,\ 
\left|\mathrm{Im}\left(\frac{V_{ub}^R}{V_{ub}^L}\right)\right|=0.551\,,\ 
|V_{ub}^L|=3.43\times 10^{-3}\,,
\label{Eq:VRBF}
\end{equation}
with $\chi^2_\text{min}/\text{dof}=2.27$. We obtain 
$\chi^2_\text{min}/\text{dof}=2.16$ in the SM and thus the scenario of 
the $b\to u$ RHC exhibits a similar consistency as the SM among the above 
experimental results in $|V_{ub}|$ determination.

As seen in Fig.~\ref{Fig:VR} and Eq.~\eqref{Eq:VRBF}, a large relative
phase between $V_{ub}^R$ and $V_{ub}^L$ is favored. We examine its 
implication for $CP$ violations in hadronic $B$ decays in the next section.

\section{\label{CPV}\boldmath $CP$ violations induced 
         by the $b\to u$ right-handed current}
As shown in the previous section, the present experimental data 
related to the $b\to u$ transition allow the right-handed $b\to u$
current with a large relative phase to the left-handed counter part.
We examine possible consequences of this new complex phase in $CP$ 
violations in hadronic $B$ decays.

\subsection{\boldmath $B\to\pi\pi$}
The time-dependent $CP$ asymmetry in $B\to\pi^+\pi^-$ is described by
the following formula \cite{BigiSanda},
\begin{equation}
A_{\pi^+\pi^-}(t)=C_{\pi^+\pi^-}\cos(\Delta m_{B_d}t)
                  -S_{\pi^+\pi^-}\sin(\Delta m_{B_d}t)\,,
\end{equation}
where the direct $CP$ asymmetry $C_{\pi^+\pi^-}$ and the mixing-induced $CP$
asymmetry $S_{\pi^+\pi^-}$ are expressed as
\begin{equation}
C_{\pi^+\pi^-}=\frac{1-|\bar\rho(\pi^+\pi^-)|^2}{1+|\bar\rho(\pi^+\pi^-)|^2}\,,
\end{equation}
and
\begin{equation}
S_{\pi^+\pi^-}=\frac{2\mathrm{Im}\left((q/p)\bar\rho(\pi^+\pi^-)\right)}
                    {1+|\bar\rho(\pi^+\pi^-)|^2}\,,
\label{Eq:CS}
\end{equation}
respectively. The amplitude ratio $\bar\rho(\pi^+\pi^-)$ is defined by
\begin{equation}
\bar\rho(\pi^+\pi^-)=\frac{A(\bar B^0\to\pi^+\pi^-)}{A(B^0\to\pi^+\pi^-)}\,,
\label{Eq:RHO}
\end{equation}
and the ratio of the $B$--$\bar B$ mixing coefficients is given as
$q/p=V_{td}^L V_{tb}^{L*}/V_{td}^{L*} V_{tb}^L$ for the $B_d$ case under
consideration here.

The isospin analysis is mandatory to extract the information on the weak 
phase in this process because of the penguin pollution \cite{GL}. The decay 
amplitudes of the isospin doublet $(B^+,B^0)$ are expressed in terms of 
the isospin amplitudes $A_I=\langle (\pi\pi)_I|B^0\rangle$ ($I=0,2$):
\begin{equation}
\label{Eq:Bpipi1}
A(B^+\to\pi^+\pi^0)=\sqrt{\frac{3}{2}}A_2\,,
\end{equation}
\begin{equation}
\label{Eq:Bpipi2}
A(B^0\to\pi^+\pi^-)=\frac{1}{\sqrt{3}}A_2+\sqrt{\frac{2}{3}}A_0\,,
\end{equation}
\begin{equation}
\label{Eq:Bpipi3}
A(B^0\to\pi^0\pi^0)=\sqrt{\frac{2}{3}}A_2-\frac{1}{\sqrt{3}}A_0\,.
\end{equation}
We note a simple triangle relation,
$A(B^+\to\pi^+\pi^0)=A(B^0\to\pi^+\pi^-)/\sqrt{2}+A(B^0\to\pi^0\pi^0)$.
The $(\bar B^0,B^-)$ decay amplitudes bear similar relations to 
$\bar A_I=\langle (\pi\pi)_I|\bar B^0\rangle$. The relative phase between 
$A_0$ and $A_2$ can be determined with a twofold ambiguity as well 
as their magnitudes by measuring the branching fractions of three decay 
modes in Eqs.~\eqref{Eq:Bpipi1}, \eqref{Eq:Bpipi2} and \eqref{Eq:Bpipi3};
and likewise for $(\bar B^0,B^-)$ and $\bar A_I$.

The ratio of $B\to\pi^+\pi^-$ amplitudes in Eq.~\eqref{Eq:RHO} is expressed
in terms of the isospin amplitudes as
\begin{equation}
\bar\rho(\pi^+\pi^-)=\frac{\bar A_2}{A_2}\frac{1+\bar z}{1+z}\,,
\end{equation}
where $z=\sqrt{2}A_0/A_2$, $\bar z=\sqrt{2}\bar A_0/\bar A_2$, and
they are obtained from the relevant branching fractions as described above.
The amplitudes of $I=2$ are determined by the tree-level $W$ boson exchange 
since the gluon penguin diagram has the nature of $\Delta I=1/2$. 
In the SM, the $I=2$ amplitudes are governed by the single weak phase of 
$V_{ub}^L$ and thus there is no $CP$ asymmetry in this channel
except the small correction due to the electroweak penguin diagrams.

In the presence of the right-handed $b\to u$ current in Eq.~\eqref{Eq:LCC},
the amplitudes of $I=2$ consist of the left- and right- handed 
contributions: $A_2=A_{2L}+A_{2R}$ and $\bar A_2=\bar A_{2L}+\bar A_{2R}$.
Thus, the large imaginary part of $V_{ub}^R/V_{ub}^L$ suggested 
by the analysis in Sec.~\ref{Sec:Vub} implies a possibility of $CP$ violation 
in the $I=2$ channel. We neglect the electroweak penguins in the following 
analysis because effects of the RHC are expected to be larger
than them.

The direct $CP$ asymmetry in $B^+\to\pi^+\pi^0$, which vanishes in the SM,
is written as
\begin{equation}\label{Eq:ACPPIPI}
 A_{CP}(B^+\to\pi^+\pi^0)=
 \frac{1-\left|R_{\pi\pi}\right|^2}{1+\left|R_{\pi\pi}\right|^2}\,,
\end{equation}
where the effect of the RHC in the $I=2$ channel is
represented by
\begin{equation}
 R_{\pi\pi}\equiv\frac{1+\bar{A}_{2R}/\bar{A}_{2L}}{1+A_{2R}/A_{2L}}\,.
\end{equation}
Other CPV observables are also affected by the RHC:
\begin{equation}
C_{\pi^+\pi^-}=
\left(1-\left|R_{\pi\pi}\right|^2 \left|\frac{1+\bar{z}}{1+z}\right|^2\right)/
\left(1+\left|R_{\pi\pi}\right|^2 \left|\frac{1+\bar{z}}{1+z}\right|^2\right)\,,
\end{equation}
\begin{equation}\label{Eq:SPIPI}
S_{\pi^+\pi^-}=\sqrt{1-C_{\pi^+\pi^-}^2}\,
\sin\left(2\phi_2^L+\arg\left(R_{\pi\pi}\right)+
\arg\left( \frac{1+\bar{z}}{1+z}\right)\right)\,,
\end{equation}
and
\begin{equation}
C_{\pi^0\pi^0}=
\left(1-\left|R_{\pi\pi}\right|^2\left|\frac{2-\bar{z}}{2-z}\right|^2\right)/
\left(1+\left|R_{\pi\pi}\right|^2\left|\frac{2-\bar{z}}{2-z}\right|^2\right)\,,
\end{equation}
where $\phi_2^L$ is one of the angles of the unitarity triangle
in Fig.~\ref{Fig:RHOETA} and $C_{\pi^0\pi^0}$ is the counter part of
$C_{\pi^+\pi^-}$ in $B\to\pi^0\pi^0$. We note that $C_{\pi^0\pi^0}$ is
determined with the time-integrated decay rate of the tagged $B\to\pi^0\pi^0$
process. The experimental data of these observables and relevant $CP$ averaged
branching fractions are summarized in Table \ref{Tab:PIPI}. The phase 
$\phi_2^L$ is extracted from the unitarity triangle construction
indicated in Fig.~\ref{Fig:RHOETA} as $\phi_2^L=84.7^\circ\pm 7.5^\circ$. 

\begin{table}
 \centering
 \begin{tabular}{ll}
  \hline
  $C_{\pi^+\pi^-}$ &
   $-0.31\pm 0.05$ \\
  $S_{\pi^+\pi^-}$ &
   $-0.66\pm 0.06$ \\
  $C_{\pi^0\pi^0}$ & 
   $-0.43\pm 0.24$ \\
  $A_{CP}(B^+\to\pi^+\pi^0)$ & 
   $-0.026\pm 0.039$ \\
  $\text{BR}(B\to\pi^+\pi^-)$ & 
   $(5.10\pm 0.19)\times 10^{-6}$ \\
  $\text{BR}(B\to\pi^0\pi^0)$ & 
   $(1.91\pm 0.225)\times 10^{-6}$ \\
  $\text{BR}(B^\pm\to\pi^\pm\pi^0)$ & 
   $(5.48\pm 0.345)\times 10^{-6}$ \\
  \hline
 \end{tabular}
 \caption{\label{Tab:PIPI}Experimental data in $B\to\pi\pi$,
          taken from the compilation by HFAG \cite{HFAG}.}
\end{table}

\begin{figure}
 \centering
 \includegraphics[width=25em]{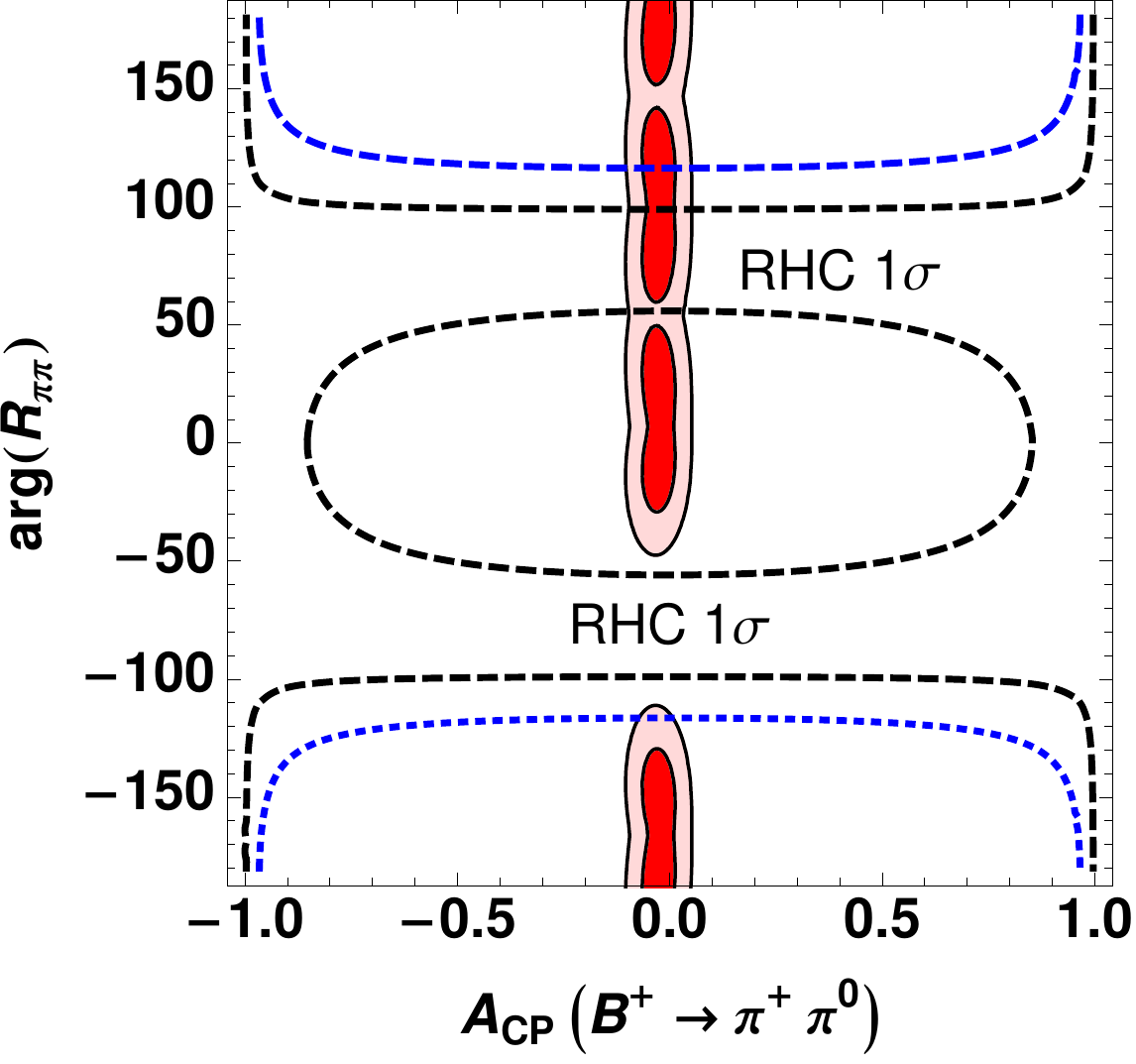}
 \caption{\label{Fig:PIPI}
  Allowed region of the direct $CP$ asymmetry and the phase discrepancy 
  in $B\to\pi\pi$. The dark (light) red region is $1\sigma$ ($2\sigma$).
  The prediction of the CPV RHC is also shown:
  The region between the black dashed ellipse and two black dashed lines 
  represents the $1\sigma$ prediction as indicated and 
  the region between the blue dotted lines is $2\sigma$.}
\end{figure}

We determine or constrain $R_{\pi\pi}$ with these data as shown in 
Fig.~\ref{Fig:PIPI}. The abscissa is the $CP$ asymmetry 
$A_{CP}(B^+\to\pi^+\pi^0)$, which is uniquely related to $|R_{\pi\pi}|$
as seen in Eq.~\eqref{Eq:ACPPIPI}, and the ordinate is 
$\arg(R_{\pi\pi})$, which represents
the possible discrepancy in the $\phi_2$ measurements between $B\to\pi\pi$ 
and the unitarity triangle as seen in Eq.~\eqref{Eq:SPIPI}.
A strong constraint is given for $A_{CP}(B^+\to\pi^+\pi^0)$, while
$\arg(R_{\pi\pi})$ is restricted rather weakly because of the eight-fold
ambiguity in the isospin analysis.

The dependence of $R_{\pi\pi}$ on $V_{ub}^R$ is obtained by
evaluating $A_{2R}/A_{2L}$ in the factorization approximation:
\begin{equation}\label{Eq:A2RA2LPIPI}
 \frac{A_{2R}}{A_{2L}}\simeq
  1.56\frac{V_{ub}^{R*}}{V_{ub}^{L*}}e^{i\delta_{\pi\pi}}\,,
\end{equation}
where we introduce a strong phase $\delta_{\pi\pi}$ as 
an arbitrary parameter, which cannot be evaluated by the factorization method. 
A similar expression is obtained for $\bar A_{2R}/\bar A_{2L}$ replacing 
$V_{ub}^{R*}/V_{ub}^{L*}$ by its complex conjugate. 
The details of the calculation using renormalization group equations (RGE) 
and the factorization is relegated in Appendix \ref{Ap:FACPIPI}. 

\begin{figure}
 \centering
 \includegraphics[width=25em]{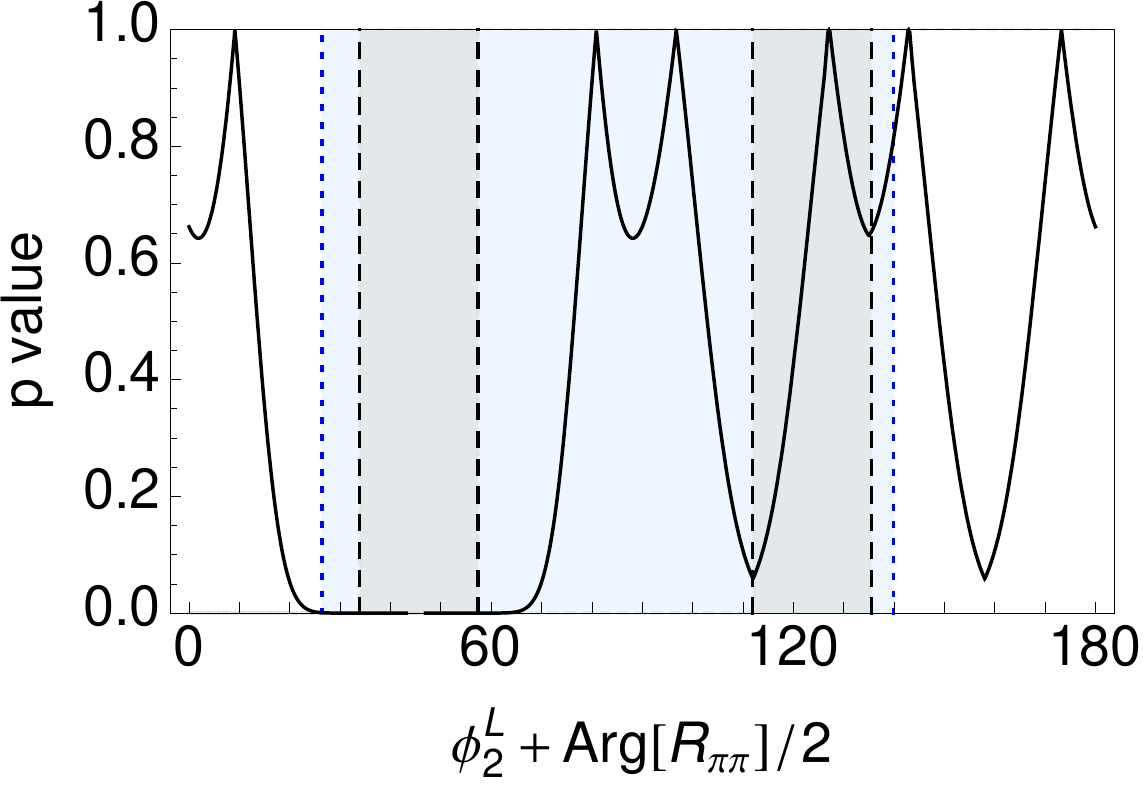}
 \caption{\label{Fig:PIPIPV}
  The $p$ value of $\phi_2^L+\arg(R_{\pi\pi})/2$ assuming 
  $\sin\delta_{\pi\pi}=0$ (solid line). The $1\sigma$ ($2\sigma$)
  prediction of the CPV RHC is also shown as the shaded region with 
  vertical black dashed (blue dotted) boundaries.}
\end{figure}

In Fig.~\ref{Fig:PIPI}, taking the strong phase $\delta_{\pi\pi}$ 
as a free parameter, we also present the prediction on 
$A_{CP}(B^+\to\pi^+\pi^0)$ and $\arg(R_{\pi\pi})$ for the allowed range 
of $V_{ub}^{R}/V_{ub}^{L}$ shown in Fig.~\ref{Fig:VR}. If 
$V_{ub}^{R}/V_{ub}^{L}$ has a significant imaginary part as suggested in 
Sec.~\ref{Sec:Vub}, the almost vanishing $A_{CP}(B^+\to\pi^+\pi^0)$ requires 
$\delta_{\pi\pi}\simeq 0$ or $\pi$, and $\arg(R_{\pi\pi})$ is sizable although
its measurement suffers from the eight-fold ambiguity mentioned above. 
Figure \ref{Fig:PIPIPV} shows the $p$ value of $\phi_2^L+\arg(R_{\pi\pi})/2$
assuming $\sin\delta_{\pi\pi}=0$ as well as its range predicted for the
allowed region of $V_{ub}^{R}/V_{ub}^{L}$ in Fig.~\ref{Fig:VR}.
The six-peak structure corresponds to the eight-fold ambiguity since 
each of the peaks at $127^\circ$ and $143^\circ$ consists of 
two solutions. The rather wide overlap between the theoretical prediction
and the experimentally allowed region is partly due to the multifold 
ambiguity, and shows that both the SM and the scenario of the CPV RHC are
consistent with the present $B\to\pi\pi$ data.

\subsection{\boldmath $B\to\rho\rho$}
The isospin analysis can be applied to $B\to\rho\rho$ as in $B\to\pi\pi$ 
provided that the helicity state of the $\rho$ mesons is identified 
by the angular analysis \cite{DQSTL}. The possible final helicity
states are $\rho_L\rho_L$ and $\rho_T\rho_T$, where $\rho_{L(T)}$ denotes
the longitudinal (transverse) helicity state of $\rho$ meson. The final state
of $\rho_T\rho_T$ is a mixture of $CP$-even and $CP$-odd states, whereas 
$\rho_L\rho_L$ is purely $CP$-even as $\pi\pi$. Hence, we can study
$CP$ violation in $B\to\rho_L\rho_L$ in a similar manner as $B\to\pi\pi$.
We note that BABAR and Belle experiments have reported the dominance of
the longitudinal final states in $B^+\to\rho^+\rho^0$ and $B\to\rho^+\rho^-$.
Although a $2.1\sigma$ difference between BABAR and Belle in the fraction of
the longitudinal state in $B\to\rho^0\rho^0$ exists \cite{BELLE:FLZZ}, 
the longitudinal fraction is likely to be sizable. 
Accordingly, we focus on $CP$ violation in $B\to\rho_L\rho_L$ in this work.

As in the above analysis of $B\to\pi\pi$, CPV observables
$C_{\rho_L^+\rho_L^-}$, $S_{\rho_L^+\rho_L^-}$, $C_{\rho_L^0\rho_L^0}$, 
$A_{CP}(B^+\to\rho_L^+\rho_L^0)$ are given in terms of $z$, $\bar z$, 
$R_{\rho_L\rho_L}$ and $\phi_2^L$.
In addition to these observables, the mixing-induced $CP$ asymmetry in
$B\to\rho_L^0\rho_L^0$, denoted as $S_{\rho_L^0\rho_L^0}$, is measurable 
and represented as
\begin{equation}
S_{\rho_L^0\rho_L^0}=\sqrt{1-C_{\rho_L^0\rho_L^0}^2}\,
\sin\left(2\phi_2^L+\arg\left(R_{\rho_L\rho_L}\right)+
\arg\left(\frac{2-\bar{z}}{2-z}\right)\right).
\end{equation}
We summarize the experimental values of the CPV parameters 
as well as the relevant branching and longitudinal fractions ($f_L$'s) 
in Table \ref{Tab:RHORHO}.

\begin{table}
 \centering
 \begin{tabular}{ll}
  \hline
  $C_{\rho_L^+\rho_L^-}$ &
   $-0.06\pm 0.13$ \\
  $S_{\rho_L^+\rho_L^-}$ &
   $-0.05\pm 0.17$ \\
  $C_{\rho_L^0\rho_L^0}$ &
   $0.2\pm 0.8\pm 0.3$ \\
  $S_{\rho_L^0\rho_L^0}$ &
   $0.3\pm 0.7\pm 0.2$ \\
  $A_{CP}(B^+\to\rho_L^+\rho_L^0)$ &
   $0.051\pm 0.054$ \\
  $\text{BR}(B^\pm\to\rho^\pm\rho^0)$ &
   $(24.0\pm 1.95)\times 10^{-6}$ \\
  $\text{BR}(B\to\rho^+\rho^-)$ &
   $(24.2\pm 3.15)\times 10^{-6}$ \\
  $\text{BR}(B\to\rho^0\rho^0)$ &
   $(0.73\pm 0.275)\times 10^{-6}$ \\
  $f_L(B^\pm\to\rho^\pm\rho^0)$ & $0.950\pm 0.016$
   \cite{BELLE:FLPZ,BABAR:FLPZ}\\
  $f_L(B\to\rho^+\rho^-)$ & $0.977\pm 0.026$ \cite{BELLE:FLPM,BABAR:FLPM}\\
  $f_L(B\to\rho^0\rho^0)$ & $0.618\pm 0.118$ \cite{BABAR:FLZZ,BELLE:FLPZ}\\
 \hline
 \end{tabular}
 \caption{\label{Tab:RHORHO}Experimental data in $B\to\rho\rho$, taken
          from HFAG \cite{HFAG} unless otherwise indicated.}
\end{table}

\begin{figure}
 \centering
 \includegraphics[width=25em]{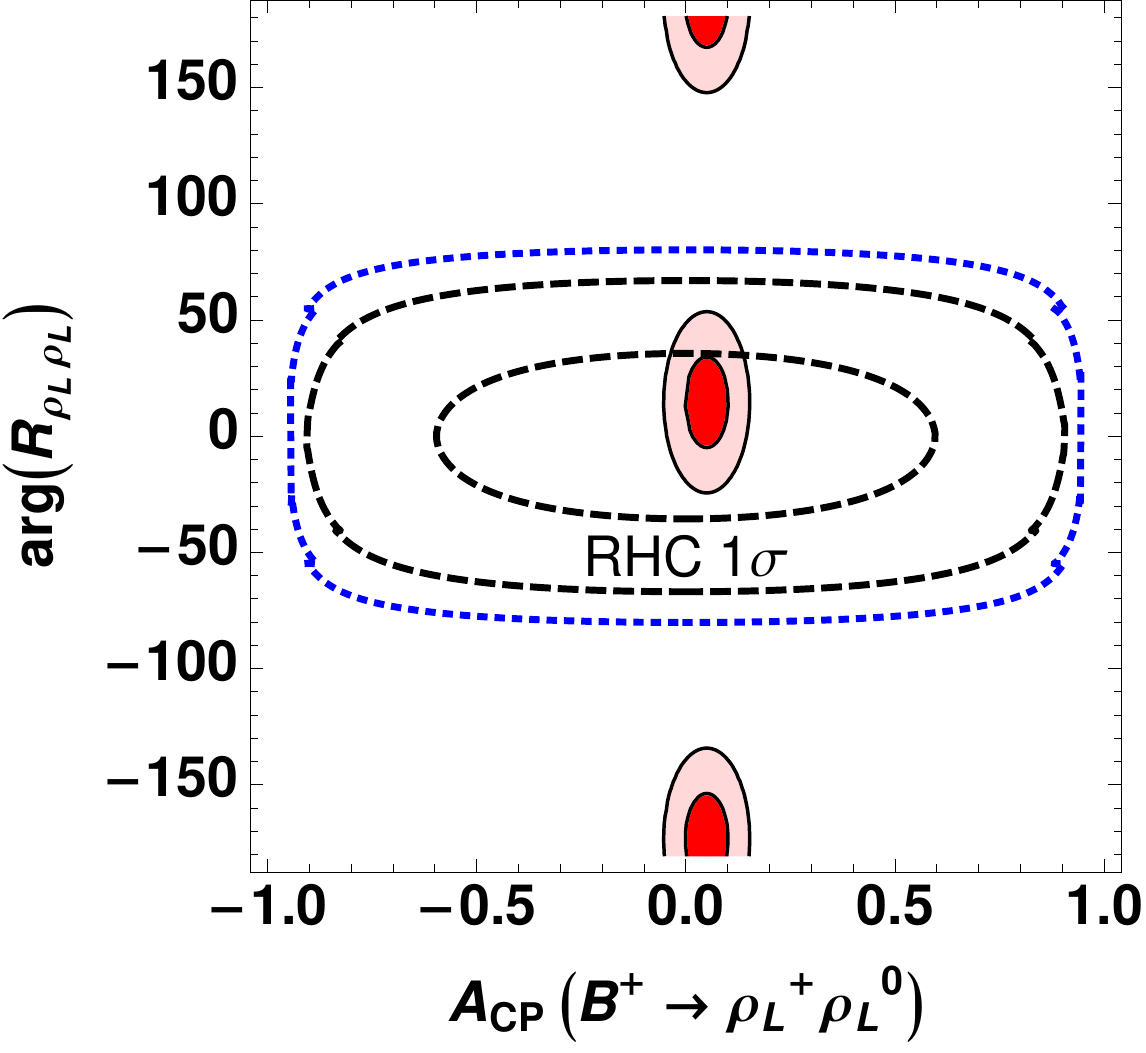}
 \caption{\label{Fig:RHORHO}
  Allowed region of the direct $CP$ asymmetry and the possible phase 
  discrepancy in $B\to\rho_L\rho_L$ is shown in the same manner as
  Fig.~\ref{Fig:PIPI}. The prediction of the CPV RHC is presented
  as well. The region between two black dashed ovals is $1\sigma$
  and the region surrounded by the blue dotted one is $2\sigma$.}
\end{figure}

These experimental data constrain $R_{\rho_L\rho_L}$ and the allowed
region is presented in Fig.~\ref{Fig:RHORHO}, in which
$A_{CP}(B^+\to\rho_L^+\rho_L^0)$ and $\arg(R_{\rho_L\rho_L})$ are 
chosen as axes. It turns out that the triangle dictated by the isospin 
relation, 
$A(B^+\to\rho_L^+\rho_L^0)
 =A(B^0\to\rho_L^+\rho_L^-)/\sqrt{2}+A(B^0\to\rho_L^0\rho_L^0)$,
and the charge-conjugated one are squashed. Hence, 
only a two-fold ambiguity remains in the isospin analysis in 
$B\to\rho_L\rho_L$ in contrast to the eight-fold one in $B\to\pi\pi$.
This reduction of the number of solutions results in a more stringent
restriction on $\arg(R_{\rho_L\rho_L})$ as seen in Fig.~\ref{Fig:RHORHO}.
In other words, the possible discrepancy in the $\phi_2$ determinations
between $B\to\rho_L\rho_L$ and the unitarity triangle are constrained 
more strongly.

\begin{figure}
 \centering
 \includegraphics[width=25em]{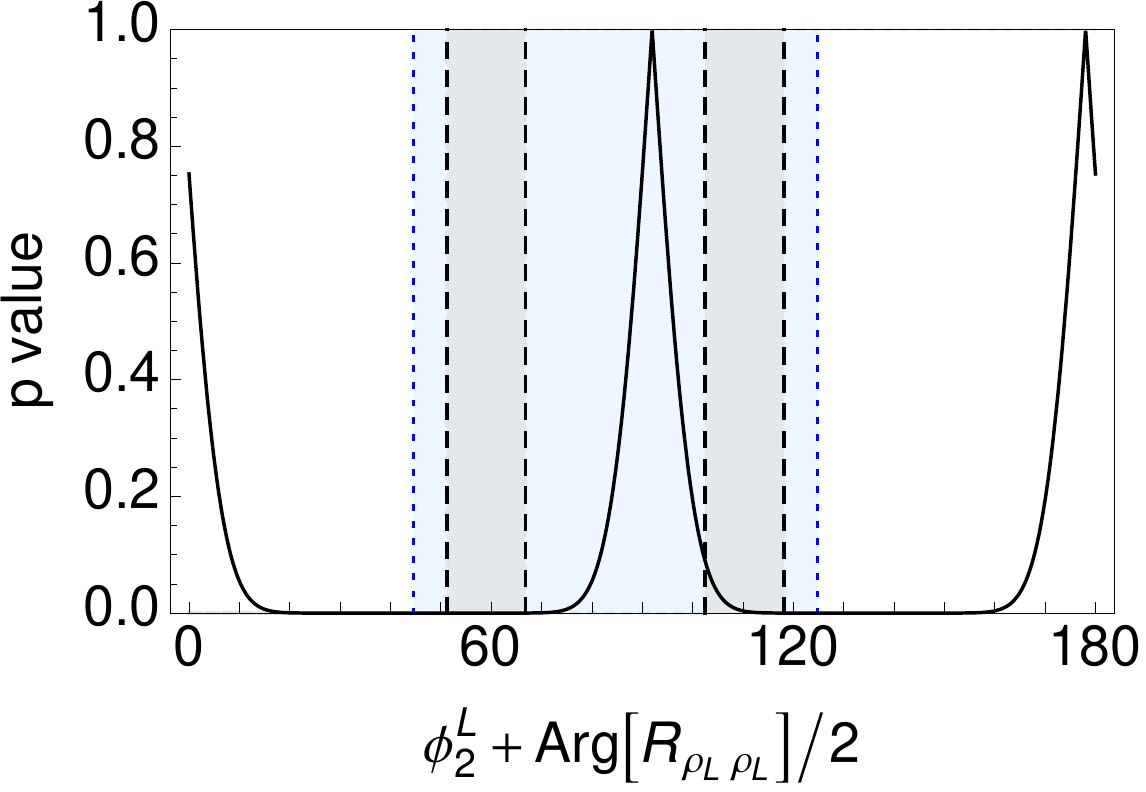}
 \caption{\label{Fig:RHORHOPV}
  The $p$ value of $\phi_2^L+\arg(R_{\rho_L\rho_L})/2$ assuming 
  $\sin\delta_{\rho_L\rho_L}=0$ and the prediction of the CPV RHC
  as in Fig.~\ref{Fig:PIPIPV}.}
\end{figure}

We evaluate $R_{\rho_L\rho_L}$ in the presence of the CPV RHC 
using the RGE and the factorization method as in the case of 
$B\to\pi\pi$. We obtain $A_{2R}/A_{2L}$ as
\begin{equation}\label{Eq:A2RA2LRHORHO}
 \frac{A_{2R}}{A_{2L}}\simeq
  -0.91\frac{V_{ub}^{R*}}{V_{ub}^{L*}}e^{i\delta_{\rho_L\rho_L}}\,,
\end{equation}
where an independent strong phase $\delta_{\rho_L\rho_L}$ is introduced.
This calculation is described in Appendix \ref{Ap:FACRHORHO}. 
The predicted region of $R_{\rho_L\rho_L}$ for the allowed 
$V_{ub}^{R}/V_{ub}^{L}$ shown in Fig.~\ref{Fig:VR} and arbitrary values
of $\delta_{\rho_L\rho_L}$ is also depicted in Fig.~\ref{Fig:RHORHO}.
One of the two experimentally allowed regions, which is consistent with 
the SM, is also compatible with the scenario of the CPV RHC.
In Fig.~\ref{Fig:RHORHOPV}, we present the $p$ value of 
$\phi_2^L+\arg(R_{\rho_L\rho_L})/2$ assuming $\sin\delta_{\rho_L\rho_L}=0$
as well as its range predicted for the allowed region of 
$V_{ub}^{R}/V_{ub}^{L}$ in Fig.~\ref{Fig:VR}. The CPV RHC is consistent 
with one of the two possible solutions that is also favored in the SM.
One may judge from Figs.~\ref{Fig:RHORHO} and \ref{Fig:RHORHOPV} that
the CPV RHC is incompatible with the experimental data at the $1\sigma$
level. However this is not the case because of the theoretical uncertainty
in the factorization. We consider that an uncertainty of a factor of 
two is likely.

\subsection{\boldmath $B\to DK$}
Two quark processes $\bar b\to \bar c u\bar s$ and $\bar b\to \bar u c \bar s$
(and their charge conjugates) give rise to $B^\pm\to DK^\pm$ decays in the SM 
and the latter is modified by the $b\to u$ RHC. We denote the relevant decay 
amplitudes in the following manner:
\begin{equation}
A(B^+\to\bar{D}^0 K^+)=A_B\,,
\end{equation}
\begin{equation}\label{Eq:DKMI2}
A(B^+\to D^0 K^+)
=A_B r_+ e^{i(\phi_{DK}+\delta_{DK})}\,,
\end{equation}
\begin{equation}
A(B^-\to D^0 K^-)=A_B\,,
\end{equation}
\begin{equation}\label{Eq:DKMI1}
A(B^-\to\bar{D}^0 K^-)
=A_B r_- e^{i(-\phi_{DK}+\delta_{DK})}\,,
\end{equation}
where amplitude ratios $r_{\pm}$ are defined to be positive.
This decay mode is employed to extract $\phi_3^L=\arg(V_{ub}^{L*})$ 
(or $\gamma$) of the unitarity triangle in the SM \cite{GLW,ADS,DPM}, 
in which the right-handed contribution vanishes and $\phi_{DK}=\phi_3^L$. 
We stress that $r_+=r_-$ in the SM, but this is not the case in the presence 
of the CPV RHC in general. Thus a direct $CP$ asymmetry,
\begin{equation}
\label{Eq:ACPDK}
A_{CP}(B^+\to D^0K^+)
=\frac{\Gamma(B^+\to D^0 K^+)-\Gamma(B^-\to \bar{D}^0 K^-)}
      {\Gamma(B^+\to D^0 K^+)+\Gamma(B^-\to \bar{D}^0 K^-)}
=\frac{r_+^2-r_-^2}{r_+^2+r_-^2}\,,
\end{equation}
is induced in addition to a discrepancy between $\phi_{DK}$ and $\phi_3^L$.
Among several methods to extract $\phi_3$ in the SM, we focus on
the most powerful one, that is the Dalitz plot method \cite{DPM}, 
in which the neutral $D$ meson in $B^\pm\to DK^\pm$ is identified with its
Dalitz decay $D\to K_S\pi^+\pi^-$. We extend the method to the case of 
$r_+\neq r_-$ in the following.

Amplitudes of the Dalitz decay are written as
\begin{align}
A(D^0\to K_S(p_K)\pi^+(p_+) \pi^-(p_-))&=A_D(s_+,s_-)\,,\\
A(\bar{D}^0\to K_S(p_K)\pi^+(p_+)\pi^-(p_-))&= A_D(s_-,s_+)\,,
\end{align}
where $s_+=(p_K+p_+)^2$ and $s_-=(p_K+p_-)^2$. We neglect small 
meson-antimeson mixing and $CP$ violation in the neutral $D$ meson system
in the present work. Then,
the differential decay rate of $B^\pm\to (K_S\pi^+\pi^-)_DK^\pm$ is
represented as
\begin{align}
&d\Gamma(B^\pm\to (K_S\pi^+\pi^-)_D K^\pm)\nonumber\\
&=|A_B|^2
  \left[|A_D(s_\mp,s_\pm)|^2+r_\pm^2|A_D(s_\pm,s_\mp)|^2
        +2r_\pm\mathrm{Re}\left\{e^{i(\pm\phi_{DK}+\delta_{DK})}
                                 A_D^*(s_\mp,s_\pm)A_D(s_\pm,s_\mp)\right\}
  \right]d\Phi\,,
\end{align}
where $d\Phi$ is a phase-space factor. 

\begin{figure}
 \centering
 \includegraphics[width=25em]{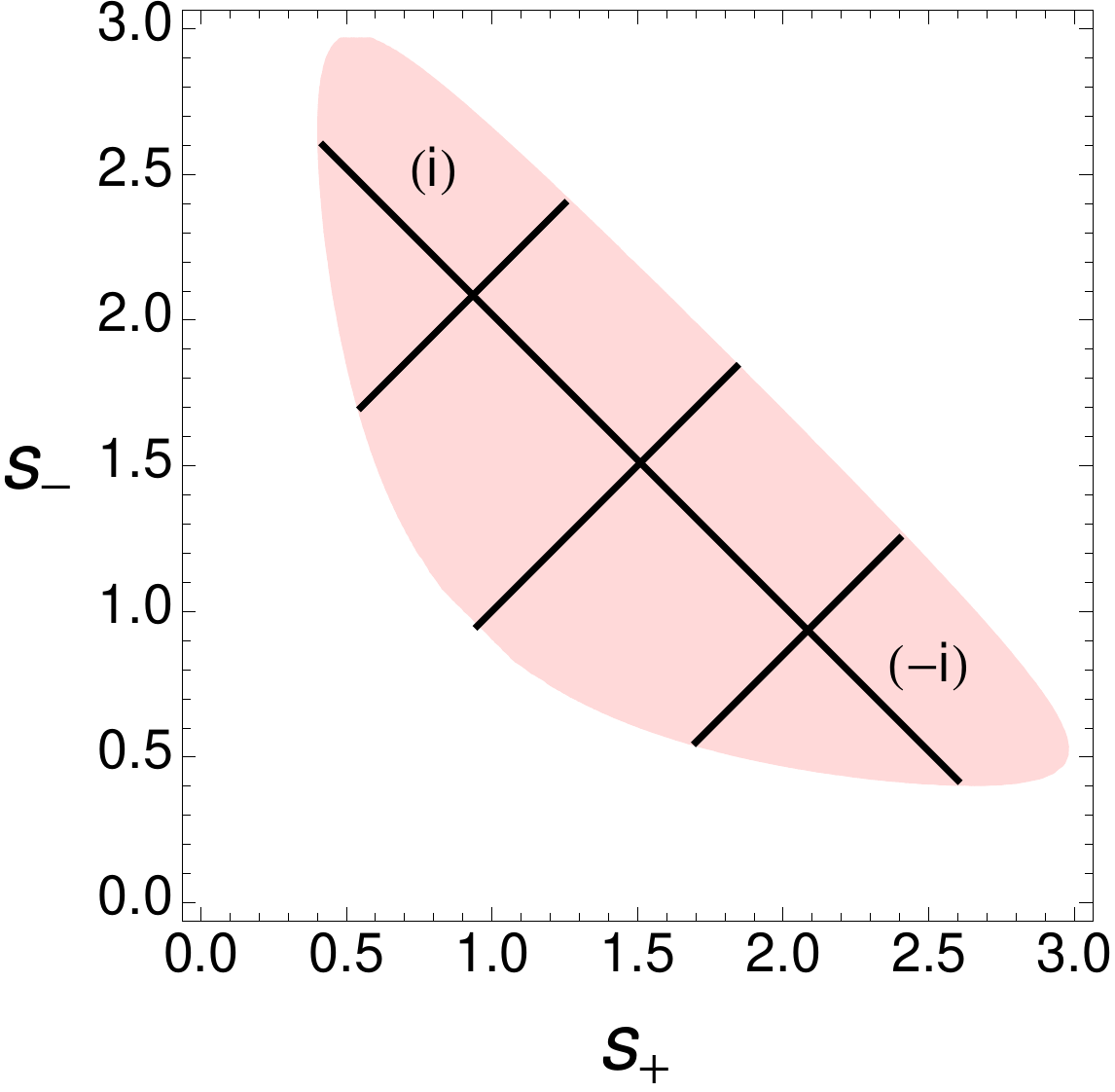}
 \caption{\label{Fig:BIN}
  An illustration of the binning in the Dalitz plot method.}
\end{figure}

In the Dalitz plot method, the phase space is divided into $2k$ bins 
as illustrated in Fig.~\ref{Fig:BIN}.
The binning is symmetric with respect to the diagonal line defined by 
$s_+=s_-$, and the $i$-th bin ($i=1,2,\cdots k$) with $s_+<s_-$
and the $(-i)$-th bin with $s_+>s_-$ form a symmetric pair.
The partial decay rate into the $i$-th bin is written as
\begin{align}
&\Gamma_i^\pm=\int_i d\Gamma(B^\pm\to(K_S\pi^+\pi^-)_D K^\pm)\nonumber\\
&=|A_B|^2\left[T_{\mp i}+r_\pm^2 T_{\pm i}+2r_\pm\sqrt{T_i T_{-i}}
               \left\{c_i\cos(\pm\phi_{DK}+\delta_{DK})
                      \mp s_i\sin(\pm\phi_{DK}+\delta_{DK})\right\}\right]\,,
\end{align}
and that into the $(-i)$-th bin is
\begin{align}
&\Gamma_{-i}^\pm=\int_{-i} d\Gamma(B^\pm\to(K_S\pi^+\pi^-)_D K^\pm)\nonumber\\
&=|A_B|^2\left[T_{\pm i}+r_\pm^2 T_{\mp i}+2r_\pm\sqrt{T_i T_{-i}}
               \left\{c_i\cos(\pm\phi_{DK}+\delta_{DK})
                      \pm s_i\sin(\pm\phi_{DK}+\delta_{DK})\right\}\right]\,,
\end{align}
where 
\begin{equation}
T_{\pm i}=\int_{\pm i}d\Phi\, |A_D(s_+,s_-)|^2\,,
\end{equation}
\begin{equation}
c_{\pm i}=\int_{\pm i}d\Phi\,
           \mathrm{Re}[A_D(s_+,s_-)A_D^*(s_-,s_+)]/
          \sqrt{T_iT_{-i}}\,,
\end{equation}
\begin{equation}
s_{\pm i}=\int_{\pm i}d\Phi\,
           \mathrm{Im}[A_D(s_+,s_-)A_D^*(s_-,s_+)]/
          \sqrt{T_iT_{-i}}\,,
\end{equation}
and we have used $c_i=c_{-i}$ and $s_i=-s_{-i}$.

The Dalitz distribution $|A_D(s_+,s_-)|^2$ is given by
the flavor-tagged neutral $D$ meson decay and thus $T_{\pm i}$'s
are known as well as $|A_B|^2$, which is determined by 
the flavor specific $D$ decay in $B^\pm\to DK^\pm$.
We notice that the number of unknown quantities ($c_i$, $s_i$, $r_\pm$, 
$\phi_{DK}$ and $\delta_{DK}$) is $2k+4$, that of observables 
($\Gamma_{\pm i}^\pm$) is $4k$, and in principle, all the unknown
quantities can be determined provided $k\geq 2$.
In particular, we can obtain the direct $CP$ asymmetry in 
Eq.~\eqref{Eq:ACPDK} and the angle discrepancy $\phi_{DK}-\phi_3^L$
with the (extended) Dalitz plot method. 

It is possible to improve the analysis by using $c_i$'s and $s_i$'s
independently extracted from data at a charm factory \cite{DPM}. 
The entangled $D^0\bar D^0$ states produced near the threshold exhibit 
quantum interference that depends on $c_i$'s and $s_i$'s. 

In Ref.~\cite{BELLE:DK}, experimental data of Belle corresponding to
$\Gamma_{\pm i}^\pm$, $T_{\pm i}$ are shown for the optimized binning 
\cite{BP} with $k=8$. The result for $c_i$'s and $s_i$'s by CLEO 
collaboration \cite{CLEO} is also summarized in Ref.~\cite{BELLE:DK}.
Using these data, we obtain a constraint on the direct $CP$ asymmetry
$A_{CP}(B^+\to D^0K^+)$ and the phase disagreement 
$\arg(R_{DK})$ ($=-2(\phi_{DK}-\phi_3^L)$, see Eq.~\eqref{Eq:PHIDK} below.)
as presented in Fig.~\ref{Fig:DK}. Although the restriction is rather
mild at present, we confirm that the extended Dalitz plot method
does work and expect a better sensitivity in future.

\begin{figure}
 \centering
 \includegraphics[width=25em]{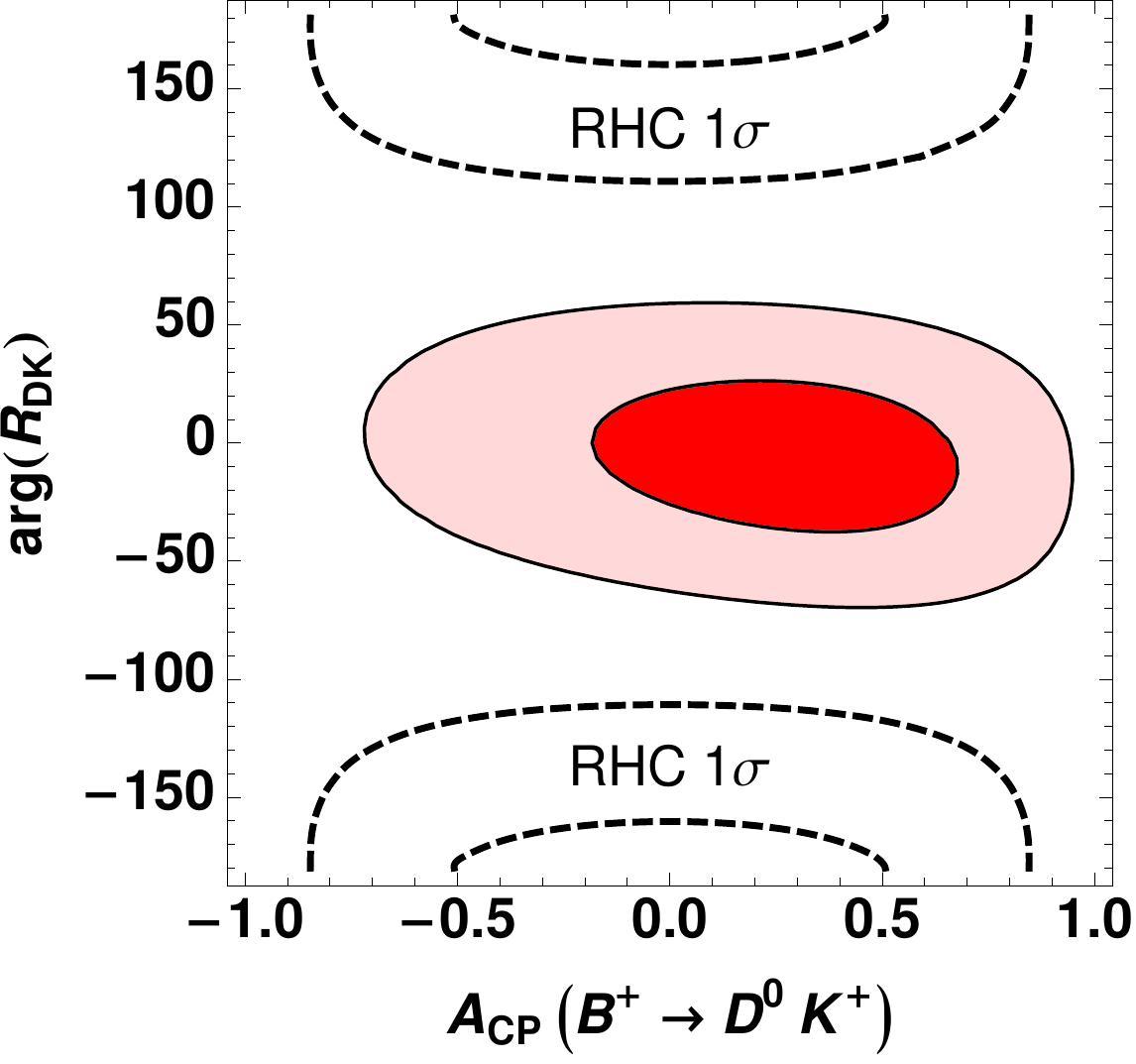}
 \caption{\label{Fig:DK}
  Allowed region of the direct $CP$ asymmetry and the possible phase 
  discrepancy in $B\to DK$ as in the same manner in Fig.~\ref{Fig:PIPI}. 
  The prediction of the CPV RHC is also shown. The region between two 
  black dashed ellipses is $1\sigma$ prediction as denoted and 
  the whole plane is practically allowed at the $2\sigma$ level.}
\end{figure}

In order for a comparison with the allowed region in Fig.~\ref{Fig:DK},
we evaluate the effect of the CPV RHC on $B^+\to D^0K^+$ and 
the charge conjugation mode. Their amplitudes are decomposed into 
the left- and right- handed contributions:
\begin{equation}
A(B^+\to D^0 K^+)=|A_L|e^{i(\phi_3^L+\delta_L)}+
                  |A_R|e^{i(\phi_3^R+\delta_R)}\,,
\end{equation}
and
\begin{equation}
A(B^-\to\bar D^0 K^-)=|A_L|e^{i(-\phi_3^L+\delta_L)}+
                      |A_R|e^{i(-\phi_3^R+\delta_R)}\,,
\end{equation}
where $\phi_3^{L(R)}=\arg(V_{ub}^{L(R)*})$ is the weak phase of the
left(right)-handed current and $\delta_{L,R}$ denote strong phases.
It is convenient to introduce an amplitude ratio as
\begin{align}
R_{DK}&=e^{2i\phi_3^L}\frac{A(B^-\to\bar D^0 K^-)}{A(B^+\to D^0 K^+)}\\
      &=\frac{1+|A_R/A_L| e^{i(-\phi_3^R+\phi_3^L+\delta)}}
             {1+|A_R/A_L| e^{i(\phi_3^R-\phi_3^L+\delta)}}\label{Eq:RDK}\,,
\end{align}
where $\delta=\delta_R-\delta_L$. Then, it is straightforward to
obtain the following relations from Eqs.~\eqref{Eq:DKMI2}, \eqref{Eq:DKMI1}
and \eqref{Eq:ACPDK}:
\begin{equation}
A_{CP}(B^+\to D^0K^+)=\frac{1-|R_{DK}|^2}{1+|R_{DK}|^2}\,,
\end{equation}
and
\begin{equation}\label{Eq:PHIDK}
\phi_{DK}=\phi_3^L-\arg(R_{DK})/2\,.
\end{equation}
The RGE and the factorization approximation gives 
\begin{equation}\label{Eq:ARALDK}
|A_R/A_L|=4.99|V_{ub}^R/V_{ub}^L|\,,
\end{equation}
as described in Appendix \ref{Ap:FACDK}. 

We evaluate $R_{DK}$ in Eq.~\eqref{Eq:RDK} for the allowed value
of $V_{ub}^R/V_{ub}^L$ shown in Fig.~\ref{Fig:VR} and $\phi_3^L$
determined by the unitarity triangle taking $\delta$ as a free parameter.
Then, we obtain theoretical prediction on $A_{CP}(B^+\to D^0K^+)$
and $\arg(R_{DK})$ as presented in Fig.~\ref{Fig:DK}.
We find that the scenario of the CPV RHC is disfavored
at the $1\sigma$ level despite the moderate current 
experimental constraint though it is not excluded at $2\sigma$.
This is due to the enhancement of the RHC contribution in the $DK$ mode 
shown in Eq.~\eqref{Eq:ARALDK} compared to those in the $\pi\pi$ and 
$\rho_L\rho_L$ modes in Eqs.~\eqref{Eq:A2RA2LPIPI} and \eqref{Eq:A2RA2LRHORHO}.
This notable sensitivity, though it is derived in the factorization
approximation, might play an important role in future experiments 
in order to probe or exclude the CPV RHC.

\subsection{\label{MSSM} Prediction of the MSSM}
It has been pointed out that the $b\to u$ RHC is induced by radiative
corrections in the MSSM \cite{CRIVELLIN2010,CN}.
The gluino-squark one-loop diagram with simultaneous insertions 
of the left-right mixing in the (3,3) component of the down-type squark 
mass matrix ($\Delta^{dLR}_{33}$) and that in the (1,3) component of
the up-type squark mass matrix ($\Delta^{uRL}_{13}$) gives the dominant
contribution and one obtains
\begin{equation}\label{Eq:SUSY}
V_{ub}^R=\frac{\alpha_s}{36\pi}\delta^{dLR}_{33}\delta^{uRL}_{13}\,,
\end{equation}
where the dimensionless mass insertion parameters are defined by 
$\delta^{dLR}_{33}=\Delta^{dLR}_{33}/M^2_\text{SUSY}$ and 
$\delta^{uRL}_{13}=\Delta^{uRL}_{13}/M^2_\text{SUSY}$, and
the masses of relevant supersymmetric partners are assumed to be
common  for simplicity and denoted by $M_\text{SUSY}$.

In Fig.~\ref{Fig:VR}, we present $V_{ub}^R/V_{ub}^L$ evaluated with
$V_{ub}^L$ in Eq.~\eqref{Eq:UT} and $V_{ub}^R$ in Eq.~\eqref{Eq:SUSY} 
for $|\delta^{dLR}_{33}\delta^{uRL}_{13}|=0.1$ and $0.3$. We observe 
that the MSSM contribution is consistent with the current experimental 
bound from $|V_{ub}|$ determination and the unitarity triangle within 
$2\sigma$ though the best fitted values do not seem to be realized. 
A future experiment like SuperKEKB/Belle II may find a signal of 
supersymmetry through the $b\to u$ RHC.

\section{\label{CONCLUSION}Conclusion}
We have studied the scenario of $b\to u$ right-handed current.
Our analysis combining the present experimental results for direct 
$|V_{ub}|$ determination with the unitarity triangle suggests
a significant CPV RHC in the $b\to u$ transition as presented in 
Fig.~\ref{Fig:VR}.

According to this analysis, we have examined CPV signals in two-body
hadronic $B$ decays: $B\to\pi\pi$, $B\to\rho\rho$ and $B\to DK$.
The expected signals in these decay modes are new direct $CP$ 
asymmetries, deviations of $\phi_2$ in $B\to\pi\pi,\rho\rho$ 
and that of $\phi_3$ in $B\to DK$; they are depicted in 
Figs.~\ref{Fig:PIPI}, \ref{Fig:RHORHO} and \ref{Fig:DK} as well as
the present experimental constraints. Although the direct $CP$ 
asymmetries in $B\to\pi\pi,\rho\rho$ are strongly constrained,
a sizable deviation of $\sim 50^\circ$ in $\phi_2$ is not excluded. 
As for $B\to DK$, the effect of RHC is enhanced by QCD radiative 
correction in the factorization approximation. Hence
the rather moderate current experimental bound tightly restricts
the CPV RHC. We have found that the consistency of the suggested CPV RHC 
with the present $B\to DK$ data is in between the $1\sigma$ and $2\sigma$
levels in the factorization approximation. 
The prediction of the MSSM is also compared to
the allowed region of $V_{ub}^R/V_{ub}^L$ as shown in Fig.~\ref{Fig:VR}.

In conclusion, the $b\to u$ right-handed current is a new physics
scenario that is still consistent with the present experimental data.
The suggested large $CP$ violation gives rise to the new $CP$ violating 
signals in hadronic $B$ decays and they may be detected in a future 
$B$ factory experiment.

\begin{acknowledgments}
We thank R.~Watanabe for discussion in the initial stage of this study.
The work of MT is supported in part by JSPS KAKENHI Grant Number 25400257.
\end{acknowledgments}

\appendix

\section{\boldmath Hadronic form factors in 
$B\to(\pi,\rho,\omega)\ell\bar\nu$}
We briefly summarize the hadronic form factors used in our numerical
analysis in the main text.

\subsection{\label{Ap:HFFP}\boldmath $B\to\pi$}
The hadronic form factor $f_+(q^2)$ in Eq.~\eqref{Eq:BpiRate} is defined as
\begin{equation}\label{Eq:BPIFF}
\langle\pi|\bar u\gamma^\mu b|\bar B\rangle
=f_+(q^2)(p_B^\mu+p_\pi^\mu)+f_-(q^2)q^\mu\,.
\end{equation}
The result of LCSR is concisely parameterized
in the following form of pole dominance \cite{BZ2005P}:
\begin{equation}
f_+(q^2)=\frac{r_1}{1-q^2/m_1^{\pi 2}}+\frac{r_2}{1-q^2/m^2_\text{fit}}\,,
\label{Eq:BZpi}
\end{equation}
where $r_1=0.744$, $m_1^\pi=5.32$ GeV, $r_2=-0.486$ and 
$m^2_\text{fit}=40.73\ \text{GeV}^2$. 

\subsection{\label{Ap:HFFV}\boldmath $B\to\rho,\omega$}
The form factors in Eqs.~\eqref{Eq:Hpm} and \eqref{Eq:Hz}
are parameterized as
\begin{equation}
A_1(q^2)=\frac{r_1^{A_1}}{1-q^2/{m_\text{fit}^{A_1}}^2}\,,
\end{equation}
\begin{equation}
A_2(q^2)=\frac{r_1^{A_2}}{1-q^2/{m_\text{fit}^{A_2}}^2}
         +\frac{r_2^{A_2}}{(1-q^2/{m_\text{fit}^{A_2}}^2)^2}\,,
\end{equation}
and
\begin{equation}
V(q^2)=\frac{r_1^V}{1-q^2/m_{1^-}^2}
       +\frac{r_2^V}{1-q^2/{m_\text{fit}^{V}}^2}\,.
\end{equation}
The LCSR gives \cite{BZ2005V} $r_1^{A_1}=0.240$, 
${m_\text{fit}^{A_1}}^2=37.51\text{GeV}^2$, $r_1^{A_2}=0.009$,
$r_2^{A_2}=0.212$, ${m_\text{fit}^{A_2}}^2=40.82\text{GeV}^2$, $r_1^V=1.045$,
$r_2^V=-0.721$, $m_{1^-}=5.32 \text{GeV}$,
${m_\text{fit}^V}^2=38.34\text{GeV}^2$ for $B\to\rho$, and
$r_1^{A_1}=0.217$, ${m_\text{fit}^{A_1}}^2=37.01\text{GeV}^2$,
$r_1^{A_2}=0.006$, $r_2^{A_2}=0.192$, 
${m_\text{fit}^{A_2}}^2=41.24\text{GeV}^2$, $r_1^V=1.006$, $r_2^V=-0.713$, 
$m_{1^-}=5.32 \text{GeV}$, ${m_\text{fit}^V}^2=37.45\text{GeV}^2$ for 
$B\to\omega$.

\section{\label{Ap:FAC} Evaluation of amplitudes by the factorization}
In this Appendix, we describe the calculation of $B\to\pi\pi,\rho\rho,DK$
amplitudes in the factorization method.

\subsection{\label{Ap:FACPIPI}\boldmath $B\to\pi\pi$}
The effective four-fermion hamiltonian that contributes to 
the $I=2$ channel in $B\to\pi\pi$ is decomposed into the left and right 
pieces as $\mathcal{H}_\text{eff}=\mathcal{H}_L+\mathcal{H}_R$, and
\begin{equation}\label{Eq:H4FXPIPI}
 \mathcal{H}_X=
  2\sqrt{2}G_F V_{ud}V_{ub}^{X*}
  \left[C_{1X}(\mu)O_{1X}(\mu)+C_{2X}(\mu)O_{2X}(\mu)\right]+\text{h.c.}\,,
\end{equation} 
where $X=L,R$ and $\mu$ denotes a renormalization scale.
The four-fermion operators are defined by
\begin{equation}\label{Eq:O1XPIPI}
O_{1X}=\bar u_L^\alpha\gamma^\nu d_L^\beta\,
       \bar b_X^\beta\gamma_\nu u_X^\alpha\,, 
\end{equation}
\begin{equation}\label{Eq:O2XPIPI}
O_{2X}=\bar u_L^\alpha\gamma^\nu d_L^\alpha\,
       \bar b_X^\beta\gamma_\nu u_X^\beta\,, 
\end{equation}
where $\alpha$ and $\beta$ are color indices. 
Wilson coefficients $C_{jX}$ ($j=1,2$) are obtained by solving 
a set of renormalization group equations in the leading order \cite{MM}.
The relevant anomalous dimensions are
\begin{equation}
\gamma_L=\frac{\alpha_s}{4\pi}
         \begin{pmatrix}
          -2 & 6 \\
           6 & -2
         \end{pmatrix}\,,\quad
\gamma_R=\frac{\alpha_s}{4\pi}
         \begin{pmatrix}
          -16 & 0 \\
           -6 & 2
         \end{pmatrix}\,,
\end{equation}
for $O_{jL}$ and $O_{jR}$ respectively.
As a result, we obtain the Wilson coefficients at the bottom quark mass scale
($m_b$):
\begin{equation}\label{Eq:WCL1}
 C_{1L}(m_b)=
 \frac{1}{2}
  \left(\left[\frac{\alpha_s(m_b)}{\alpha_s(m_W)}\right]^{-6/23}-
         \left[\frac{\alpha_s(m_b)}{\alpha_s(m_W)}\right]^{12/23}
  \right)\simeq -0.27\,,
\end{equation}
\begin{equation}\label{Eq:WCL2}
 C_{2L}(m_b)=
 \frac{1}{2}
 \left\{\left[\frac{\alpha_s(m_b)}{\alpha_s(m_W)}\right]^{-6/23}+
        \left[\frac{\alpha_s(m_b)}{\alpha_s(m_W)}\right]^{12/23}
 \right\}\simeq 1.12\,,
\end{equation}
\begin{equation}\label{Eq:WCR1}
 C_{1R}(m_b)=
 \frac{1}{3}
 \left\{\left[\frac{\alpha_s(m_b)}{\alpha_s(m_W)}\right]^{24/23}-
        \left[\frac{\alpha_s(m_b)}{\alpha_s(m_W)}\right]^{-3/23}
 \right\}\simeq 0.34\,,
\end{equation}
and
\begin{equation}\label{Eq:WCR2}
 C_{2R}(m_b)=\left[\frac{\alpha_s(m_b)}{\alpha_s(m_W)}\right]^{-3/23}
            \simeq 0.92\,,
\end{equation}
where $\alpha_s(m_Z)=0.118$ \cite{PDG} and $m_b=4.2\ \text{GeV}$ 
\cite{XZZ} are used. We neglect the gluon penguin operators since 
they do not contribute to the $I=2$ final state.

The amplitude ratio $A_{2R}/A_{2L}$ is conveniently evaluated
by calculating $B^+\to\pi^+\pi^0$ amplitudes,
$\langle\pi^+\pi^0|\mathcal{H}_X|B^+\rangle$, at $\mu=m_b$:
\begin{align}
\langle\pi^+\pi^0|\mathcal{H}_L|B^+\rangle
\simeq&\frac{G_F}{\sqrt{2}}V_{ud}V_{ub}^{L*}
       \left[\{C_{1L}(m_b)+C_{2L}(m_b)/3\}
         \langle\pi^0|\bar u\gamma^\nu\gamma_5 u|0\rangle
         \langle\pi^+|\bar b\gamma_\nu d|B^+\rangle\right.\nonumber\\
      &\left.+\{C_{2L}(m_b)+C_{1L}(m_b)/3\}
         \langle\pi^+|\bar u\gamma^\nu\gamma_5 d|0\rangle
         \langle\pi^0|\bar b\gamma_\nu u|B^+\rangle\right]\,,
        \label{Eq:PIPIFACL}
\end{align}
and
\begin{align}
\langle\pi^+\pi^0|\mathcal{H}_R|B^+\rangle
\simeq&\frac{G_F}{\sqrt{2}}V_{ud}V_{ub}^{R*}
       \left[2\{C_{1R}(m_b)+C_{2R}(m_b)/3\}
         \langle\pi^0|\bar u\gamma_5 u|0\rangle
         \langle\pi^+|\bar bd|B^+\rangle\right.\nonumber\\
      &\left.+\{C_{2R}(m_b)+C_{1R}(m_b)/3\}
         \langle\pi^+|\bar u\gamma^\nu\gamma_5 d|0\rangle
         \langle\pi^0|\bar b\gamma_\nu u|B^+\rangle\right]\,,
        \label{Eq:PIPIFACR}
\end{align}
where we have used Fierz rearrangement and ignored annihilation terms
and nonfactorizable contributions. The matrix elements of the vector currents
in Eqs.~\eqref{Eq:PIPIFACL} and \eqref{Eq:PIPIFACR} are
expressed by the form factors in Eq.~\eqref{Eq:BPIFF} and
those of the axial-vector currents are given by the pion decay constant.
The (pseudo)scalar operator in Eq.~\eqref{Eq:PIPIFACR}
are related to the corresponding (axial-)vector operator using the equation
of motion of the quark fields and thus its matrix element is also written 
in terms of the form factors (the decay constant). Interestingly,
we do not need to specify the values of the form factors and the decay 
constant since they disappear in the ratio 
$\langle\pi^+\pi^0|\mathcal{H}_R|B^+\rangle/
 \langle\pi^+\pi^0|\mathcal{H}_L|B^+\rangle$. 
Hence we obtain
\begin{align}\label{Eq:A2RA2LPIPIAP}
&\frac{A_{2R}}{A_{2L}}=\frac{\langle\pi^+\pi^0|\mathcal{H}_R|B^+\rangle}
                           {\langle\pi^+\pi^0|\mathcal{H}_L|B^+\rangle}
 \nonumber\\
&\simeq\frac{V_{ub}^{R*}}{V_{ub}^{L*}}\,\frac{3}{4}
       \left[\frac{C_{2R}(m_b)+C_{1R}(m_b)/3}{C_{2L}(m_b)+C_{1L}(m_b)}
             +\frac{C_{1R}(m_b)+C_{2R}(m_b)/3}{C_{2L}(m_b)+C_{1L}(m_b)}
              \frac{m_\pi^2}{m_q M_b}\right]
 \simeq 1.56\frac{V_{ub}^{R*}}{V_{ub}^{L*}}\,,
\end{align}
where $M_b$ denotes the bottom quark pole mass, $m_q$ represents
the average current mass of the up and down quarks, and we employ
$M_b=4.91\ \text{GeV}$ \cite{XZZ} and $m_q=3.5\ \text{MeV}$ \cite{PDG}
in our numerical calculation.

The factorization method described above should be understood as
a crude approximation that provides an order of magnitude.
We consider that an uncertainty of a factor of two remains 
even in the ratio in Eq.~\eqref{Eq:A2RA2LPIPIAP}.
Furthermore it gives no information on the phase shift by the strong
interaction, and thus we introduce a strong phase in 
Eq.~\eqref{Eq:A2RA2LPIPI} by hand.

\subsection{\label{Ap:FACRHORHO}\boldmath $B\to\rho\rho$}
The effective four-fermion hamiltonian for $B\to\pi\pi$,
shown in Eqs.~\eqref{Eq:H4FXPIPI}, \eqref{Eq:O1XPIPI} and \eqref{Eq:O2XPIPI},
also describes the $I=2$ amplitudes in $B\to\rho_L\rho_L$.
The relevant matrix elements are evaluated almost in the same way as 
in the case of $B\to\pi\pi$. We finally obtain
\begin{equation}
\frac{A_{2R}}{A_{2L}}=\frac{\langle\rho_L^+\rho_L^0|\mathcal{H}_R|B^+\rangle}
                           {\langle\rho_L^+\rho_L^0|\mathcal{H}_L|B^+\rangle}
 =-\frac{V_{ub}^{R*}}{V_{ub}^{L*}}\,\frac{3}{4}
  \frac{C_{2R}(m_b)+C_{1R}(m_b)/3}{C_{2L}(m_b)+C_{1L}(m_b)}
 \simeq -0.91\frac{V_{ub}^{R*}}{V_{ub}^{L*}}\,.
\end{equation}

\subsection{\label{Ap:FACDK}\boldmath $B\to DK$}
The effective hamiltonian for $B^+\to D^0 K^+$ and its charge conjugation
is given by
$\mathcal{H}_\text{eff}=\mathcal{H}_L+\mathcal{H}_R$ and
\begin{equation}\label{Eq:H4FXDK}
 \mathcal{H}_X=
  2\sqrt{2}G_F V_{cs}V_{ub}^{X*}
  \left[C_{1X}(\mu)O_{1X}(\mu)+C_{2X}(\mu)O_{2X}(\mu)\right]+\text{h.c.}\,,
\end{equation} 
where the four-fermion operators are defined by
\begin{equation}\label{Eq:O1XDK}
O_{1X}=\bar c_L^\alpha\gamma^\nu s_L^\beta\,
       \bar b_X^\beta\gamma_\nu u_X^\alpha\,, 
\end{equation}
\begin{equation}\label{Eq:O2XDK}
O_{2X}=\bar c_L^\alpha\gamma^\nu s_L^\alpha\,
       \bar b_X^\beta\gamma_\nu u_X^\beta\,.
\end{equation}
The renormalization of these operators are the same as those in 
$B\to\pi\pi$ in the leading order and hence the Wilson coefficients are 
also given by Eqs.~\eqref{Eq:WCL1}, \eqref{Eq:WCL2}, \eqref{Eq:WCR1} and
\eqref{Eq:WCR2}.

Using Fierz rearrangement and ignoring annihilation terms and 
nonfactorizable contributions, we evaluate the amplitude ratio $A_R/A_L$ 
as in the case of $B\to\pi\pi$. Eventually, we obtain 
\begin{equation}\label{Eq:ARALDK2}
\left|\frac{A_{R}}{A_{L}}\right|
 =\left|\frac{V_{ub}^{R*}}{V_{ub}^{L*}}\right|\,\frac{2 m_D^2}{M_b M_c}
  \frac{C_{2R}(m_b)+3 C_{1R}(m_b)}{C_{2L}(m_b)+3 C_{1L}(m_b)}
 \simeq 4.99\left|\frac{V_{ub}^{R*}}{V_{ub}^{L*}}\right|\,,
\end{equation}
where $M_c=1.77\ \text{GeV}$ denotes the charm quark pole mass \cite{XZZ} 
and we have neglected the up and strange quark masses.
The quark pole masses emerge when we utilize the equations of motion of 
the quark fields in order to evaluate the contribution of the RHC. 
We note that the $B\to K$ form factors and the $D$ meson decay constant 
appearing in each amplitude $A_X$ cancel out in the ratio of 
Eq.~\eqref{Eq:ARALDK2} as in $B\to\pi\pi$.

\end{document}